\documentclass[aip,reprint]{revtex4-1}

\usepackage{graphicx}
\usepackage{dcolumn}
\usepackage{bm}

\usepackage[utf8]{inputenc}
\usepackage[T1]{fontenc}
\usepackage{mathptmx}
\usepackage{gensymb}
\usepackage{siunitx}

\begin{document}

\title{An antipodal Vivaldi antenna for improved far-field properties and polarization manipulation of broadband terahertz quantum cascade lasers}

\author{U. Senica}
 \email{usenica@phys.ethz.ch}
\author{E. Mavrona}%
\altaffiliation[Now with ]{EMPA, Switzerland}
\author{T. Olariu}
\author{A. Forrer}
\author{M. Shahmohammadi}
\author{M. Beck}
\author{J. Faist}
\author{G. Scalari}
 \email{scalari@phys.ethz.ch}

\affiliation{ 
Institute for Quantum Electronics, ETH Zurich, Auguste-Piccard-Hof 1, 8093 Zurich, Switzerland
}%


\date{\today}

\begin{abstract}
We present an antipodal Vivaldi antenna for broadband double metal waveguide terahertz quantum cascade lasers and frequency combs. Its exponentially curved flare profile results in an adiabatic in-plane mode expansion, producing an improved far-field with a single-lobed beam of (23\degree$\times$19\degree) full width half maximum with octave-spanning bandwidth. The antenna acts also as a wave retarder, rotating the polarization from vertical towards horizontal polarization by a frequency-dependent angle. The laser's emission spectrum and current-voltage characteristics are not affected, as well as frequency comb operation.  Measurements agree well with numerical simulations, and the proposed antenna covers a broad spectral range (1.5-4.5 THz). 
\end{abstract}

\maketitle

Terahertz quantum cascade lasers (THz QCLs)\cite{KohlerNat2002} have been of large interest in the context of very broad emission spectrum and frequency comb operation \cite{BurghoffNatPhot2014,Roesch2014}. Double metal waveguide-based THz QCLs \cite{williamsAPL2003} have excellent properties in terms of field confinement, large bandwidth and high temperature operation \cite{BoscoAPL2019}. However, as the two metallic plates confine the optical mode to subwavelength dimensions, the waveguide's end facet acts almost as a point-like source of radiation, resulting in a highly divergent far-field. There have been several approaches to improve this drawback. For single mode lasers, the last 15 years have seen the development of several interesting solutions, from distributed feedback lasers operating on 3rd order Bragg condition \cite{amanti2009} to 1D photonic heterostructures \cite{xu2012,xu2014}, 2D ring DFBs \cite{liang2013}, 2D photonic crystal lasers \cite{chassagneux2009,chassagneux2010,halioua2014}, photonic quasi-crystals \cite{vitiello2014}, metasurface THz VECSELs \cite{CurwenNatPhot2019}, integrated patch antennae \cite{bonzon2014patch,bosco2016,justen2016}, plasmonic lasers \cite{wu2016}, and random lasers \cite{schoenhuber2016}. For frequency comb applications, all these solutions are not easy to be extended to broadband operation. Instead, lenses can be abutted to the laser's facet  \cite{lee2007lens, Wan2018}, or a horn antenna can be attached to the laser ridge - these solutions require invasive post-processing fabrication steps and non-planar technology\cite{amanti2007,MaineultAPL2008HORN}. Integrated solutions derived from previously mentioned 3rd order DFBs \cite{amanti2009} displayed good results but still with a relatively limited bandwidth\cite{RoeschExtractAPL2017}. The aim of this work is to demonstrate a broadband extractor/collimator which does not affect the laser performance allowing comb operation and which improves the far-field properties over octave-spanning spectral bandwidths. 
\par
We developed a broadband, narrow beam Vivaldi antenna. Originally demonstrated in the GHz range, the Vivaldi antenna is a member of aperiodic continuously scaled antenna structures \cite{GibsonVivaldi1979}. In general, its radiation is produced in a non-resonant fashion by travelling waves along the curved antenna shape. Over four decades of development resulted in a plethora of various shapes and subtypes, establishing Vivaldi antennas as broadband solutions for diverse applications in the microwave region \cite{VivaldIEEE2010, Yang2008, Reid2012, He2014}. 
\par 
Adapting the concept for THz QCLs, we use an antipodal Vivaldi antenna design \cite{Hood2008} which consists of two exponentially curved metallic flares extending from the end facet of a double metal waveguide. Once the optical mode propagating inside the waveguide reaches the end facet, it remains confined between the two antenna flares. With increasing distance from the waveguide, the separation between the antenna flares slowly increases due to their curved shape. The optical mode follows this increasing separation and gets adiabatically expanded, producing a narrow beam far-field pattern. As mentioned, the radiation mechanism is not based on a narrow resonance and is inherently broadband. Additionally, the antenna has an effect on the polarization. While the electric field is initially purely vertically polarized, as required by selection rules for intersubband gain in the vertical stack of quantum wells, it gets rotated while propagating through the antenna. 

\begin{figure*}
\includegraphics[width=1\textwidth]{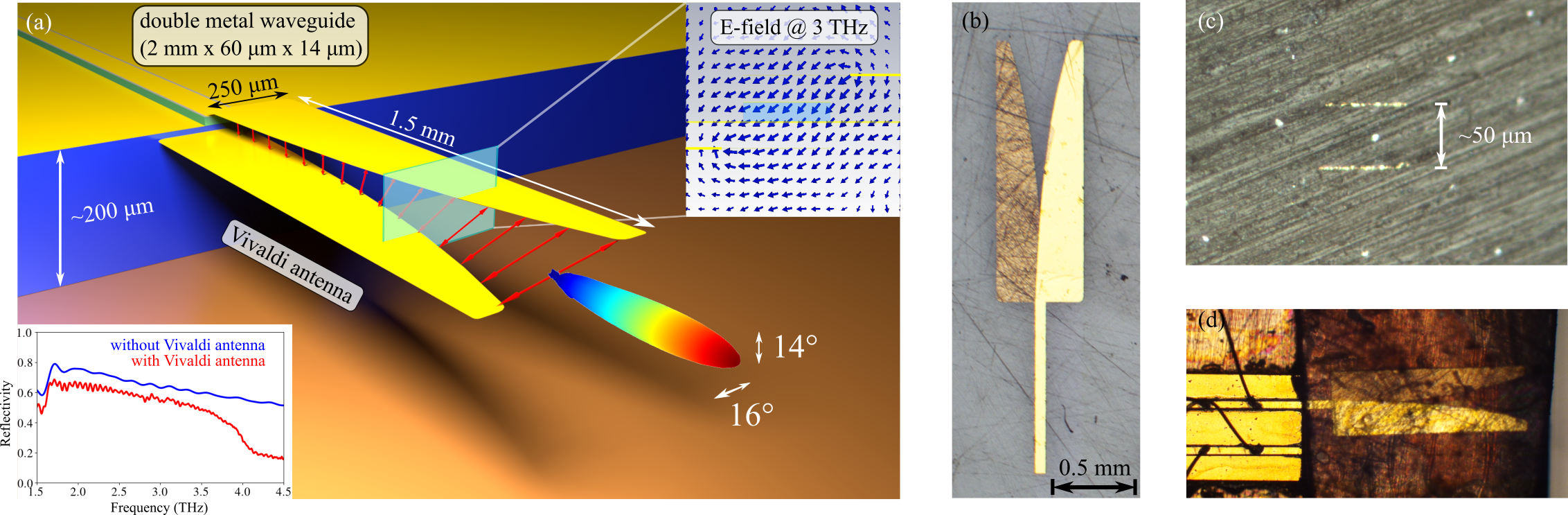}
\caption{\label{fig:sim_farfield}  (a) Illustration of an antipodal Vivaldi antenna, attached in front of a double metal waveguide THz QCL. A spherical plot of the simulated far-field of a broadband device with a flat emission spectrum between 1.5-4.5 THz is superimposed to the structure, with a FWHM of (16\degree$\times$14\degree). The red arrows represent the electric field vectors of propagating light waves, which follow the antenna's shape. Top right inset shows the simulated electric field vectors in the cross-section of the antenna at a distance of $\simeq$1 mm from the waveguide at a frequency of 3 THz (see online version for an animation) (Multimedia view). In the bottom left inset is the computed reflectivity for a double metal waveguide with and without a Vivaldi antenna. The slight drop in reflectivity is mainly due to the polymer between the antenna flakes which has a refractive index of $n\simeq1.53$. (b) A processed Vivaldi antenna. The double-sided structure is fabricated on a low-loss transparent polymer using standard photolithography, metalization and lift-off techniques. (c) Front view of a polished interface surface of a tri-layer topas stack with the antenna structure in the middle layer. (d) Top view of a Vivaldi antenna mounted in front of a double metal waveguide THz QCL.}
\end{figure*}

\par 
To obtain a suitable design, initially an example from microwaves \cite{Gazit1988} was downscaled to THz frequencies. A series of full-wave 3D numerical simulations using CST Microwave Studio\textsuperscript{®} was performed, where a broadband time-domain pulse was launched from the waveguide into the antenna, and the resulting far-field was computed in the frequency range of interest (1.5 - 4.5 THz). The initial field was the fundamental propagating waveguide mode, calculated using an eigenmode solver in a 2D cross-section of the waveguide. Initially, we studied the effects of the various geometric parameters, i.e., the antenna width, length, shape of the opening, position and separation between the two parts, on the performance. The total antenna length and the width of the opening turned out to be the dominant parameters. After a parametric sweep of these two parameters, the final optimized design was obtained. The predicted beam width is (16\degree$\times$14\degree) full width half maximum (FWHM) for a device with a flat emission spectrum covering two octaves in frequency (1.5 - 4.5 THz), see Fig. \ref{fig:sim_farfield}.(a).

\par 
The antipodal Vivaldi antenna design features two independent antenna arms separated by a certain distance, requiring a hosting substrate. We used cyclic olefin copolymer (TOPAS\textsuperscript{®} COC), a transparent polymer with low losses in the THz region\cite{Cunningham2011} and a refractive index of $n\simeq1.53$ (the polymer was also included in the far-field simulations described in the previous paragraph). For the fabrication, the shape of the antenna flare was transferred to the polymer using standard photolithography. Following a metalization and lift-off step, an array of half-antennas was made. Since TOPAS is transparent at visible wavelengths, the sample was flipped upside down and the photolithography mask could be aligned to the bottom antenna layer with relative ease. Repeating the same photolithography, metalization and lift-off steps as before, the full antenna array was completed. An optical microscope image of a fabricated Vivaldi antenna is shown in Fig. \ref{fig:sim_farfield}.(b).
\par 
For mounting, an individual antenna was removed from the array by cutting through any point of a straight waveguide-matching section (simulations predict that this additional section does not deteriorate the far-field). Special care had to be taken to mount it in good alignment with the waveguide. The latter generally has a thickness between 8 - 14 \SI{}{\micro\metre} (in our case 14 \SI{}{\micro\metre}) and is attached on top of a carrier substrate with a typical thickness of $d_\mathrm{{sub}}\approx200$ \SI{}{\micro\metre}. To ensure a matching vertical position of the waveguide and antenna, a tri-layer polymer stack was prepared using thermocompression wafer bonding. The antenna layer ($d_\mathrm{{ant}}\approx50$ \SI{}{\micro\metre}) was encapsulated between two topas layers with $d_\mathrm{{top}}\approx (d_\mathrm{{sub}}-15$ \SI{}{\micro\metre}$)$, leaving some margin for eventual mounting imprecisions. After polishing the front/back surfaces with a diamond sheet (shown in Fig. \ref{fig:sim_farfield}.(c)), the antenna stack was placed in front of a double metal waveguide THz QCL. The horizontal alignment was adjusted using a mechanical stage, and the position was fixed using UV glue, as shown in Fig. \ref{fig:sim_farfield}.(d). Eventual contact with the laser facet is not resulting in damages (as frequently observed with lenses) due to the relatively soft nature of TOPAS.

\begin{figure*}
\includegraphics[width=0.8\textwidth]{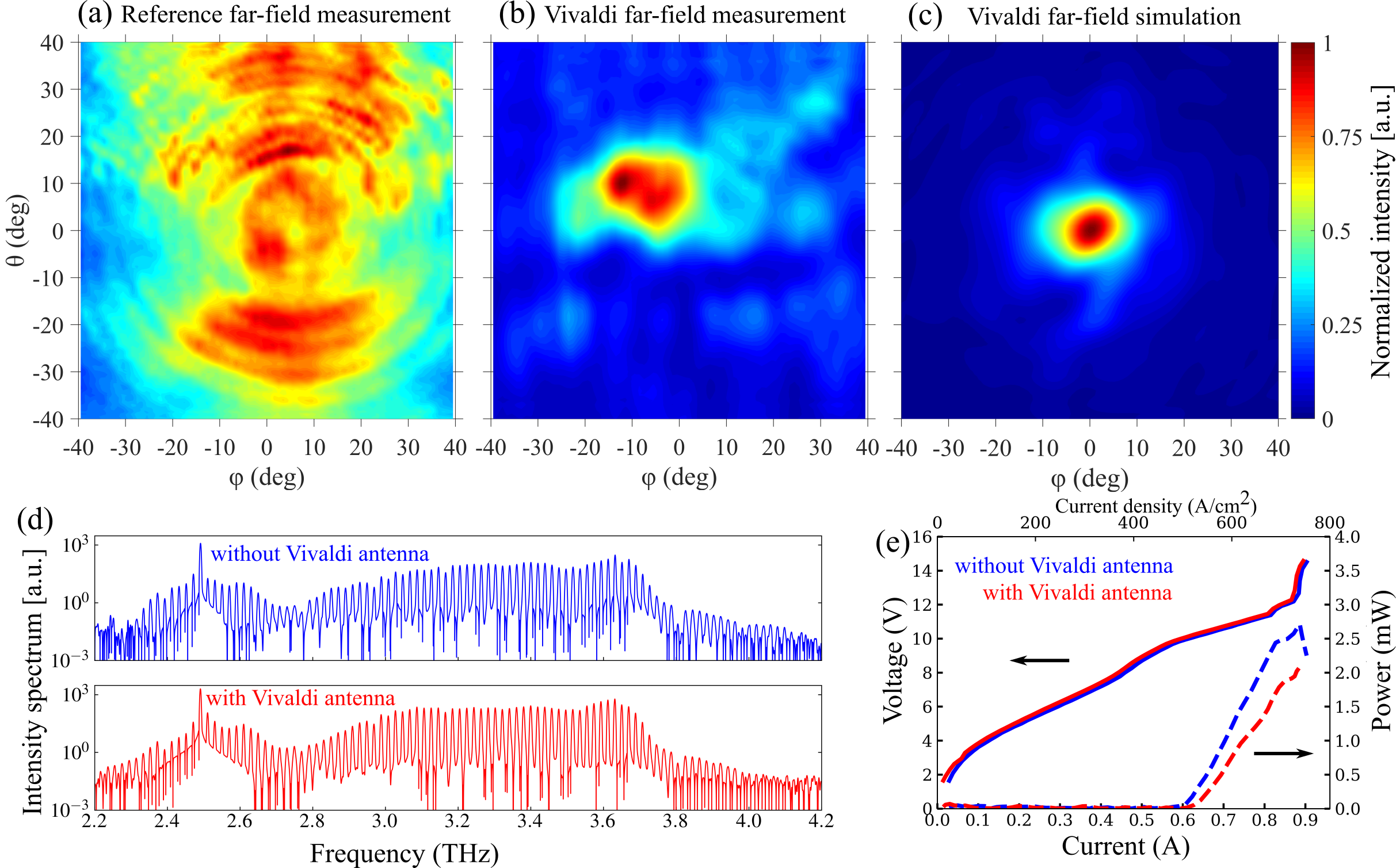}
\caption{\label{fig:meas_farfield_spec_liv} (a) Reference far-field measurement of a broadband THz QCL, where the intensity is spread out to a wide range of angles. (b) Measured far-field of a broadband THz QCL with a mounted Vivaldi antenna, with a FWHM of (23\degree$\times$19\degree). (c) Simulated far-field with the Vivaldi antenna has a FWHM of (16\degree$\times$14\degree). The computed frequency-dependent far-field contributions were normalized using the emission spectrum of the measured device.  (d) Measured emission spectrum of the THz QCL before and after mounting the Vivaldi antenna. The broad emission spans between approximately 2.3-4.0 THz and does not change significantly after mounting the antenna. (e) Light-current-voltage (L-I-V) curves before and after mounting the Vivaldi antenna in pulsed mode. The IV characteristics overlap, while the total collected power decreases by $\approx$20\% due to absorption in the polymer (measured losses in ref. \cite{Cunningham2011} were  $\approx$1-2 cm\textsuperscript{-1}) and gold. The measurements were done in pulsed mode at a temperature of 15 K. In (a), the pulse length was 1 \SI{}{\micro\second} with a duty cycle of 20\%. In the other measurements, the pulse length was 200 ns at 30\% duty cycle in (b-d), and 10\% duty cycle in (e).}
\end{figure*}

\par 
The far-field patterns of the fabricated devices were measured using a pyroelectric detector (Gentec-EO: THZ2I-BL-BNC) mounted on a motorized scanning stage. The laser was driven in micro-pulse (200 ns pulses, 30\% duty cycle), macro-pulse mode (30 Hz, 50\% duty cycle). First, we measured a reference far-field of a broadband THz QCL (same epilayer as in ref. \cite{RoeschNanophotonics2018}) with a standard mounting on the edge of a copper submount. As shown in Fig. \ref{fig:meas_farfield_spec_liv}.(a), the far-field intensity is spread over a wide range of angles, supporting predictions by simulations. In contrast, measuring a broadband THz QCL with a mounted Vivaldi antenna results in a single-lobed far-field pattern with a FWHM of (23\degree$\times$19\degree), see Fig. \ref{fig:meas_farfield_spec_liv}.(b). The offset of the main lobe in the measured far-field is a consequence of a slight misalignment of the center of the far-field stage with the QCL/antenna axis. Intensity outside the main lobe is detected due to a non-perfect interface and alignment between the waveguide and antenna. The measured far-field is in good agreement with full-wave 3D simulation results shown in Fig. \ref{fig:meas_farfield_spec_liv}.(c), with a predicted beam width of (16\degree$\times$14\degree) FWHM. To make the comparison more realistic, we used the measured emission spectrum from Fig. \ref{fig:meas_farfield_spec_liv}.(d) in the computation of the simulated far-field. First, we approximated the spectrum as a set of discrete modes with a separation of $\Delta f = 20$ GHz, since this is close to the actual mode separation in the measured device with a length of 2 mm. The relative amplitude of each discrete mode was calculated by integrating the spectrum intensity in each frequency window:
\begin{equation}
    I(f_n) = \int_{f_n-\Delta f/2}^{f_n+\Delta f/2}\ I(f) \ df, \ \ \Delta f = 20\  \mathrm{GHz} 
\end{equation}
Then we computed the simulated far-field at each discrete frequency $F(f_n)$. The final broadband computed far-field is then a linear superposition of individual far-fields, normalized by the spectral intensity of each discrete mode:
\begin{equation}
    F_{\mathrm{\ broad}} =  \sum_{n=1}^{N} I(f_n) \times F(f_n)
\end{equation}

\par 
We measured the emission spectrum of the THz QCL before and after mounting the Vivaldi antenna in pulsed mode (200 ns pulses, 30\% duty cycle). As shown in Fig. \ref{fig:meas_farfield_spec_liv}.(d), the emission spectrum is very broad, spanning between approximately 2.3-4.0 THz, and does not change significantly after mounting the Vivaldi antenna. This confirms that the antenna works well with broadband devices without having an effect on the emission spectrum. Very similar results in terms of far-field, light-voltage-current (L-I-V) curves and spectra were measured on another laser device equipped with another Vivaldi antenna.
\par 
A comparison of the L-I-V curves in pulsed mode (200 ns pulses, 10\% duty cycle at 15 K) in Fig. \ref{fig:meas_farfield_spec_liv}.(e) indicates that the Vivaldi antenna does not have an effect on the current-voltage characteristics. The measured peak power is reduced from around 2.7 mW to 2.1 mW ($\approx$20\%) for a 10\% duty cycle. This measurement was done with a broad area absolute THz Power Meter by Thomas Keating Ltd (TK), placed directly outside the cryostat to collect all of the power coming through the window (including the unfocused reference beam). Since the IV curve does not change, the total emitted power does not change either, and the loss in collected power is due to Fresnel reflection and propagation losses in the polymer and gold in the Vivaldi antenna. However, despite a slight decrease in total power, the power areal density is increased due to the improved far-field as compared to a standard double metal waveguide. This is supported by far-field measurements: the maximum measured signal was increased by 48\%, and an integral of the measured detector intensity over the 2D FWHM beam area (a circle with a diameter of 21\degree) was larger by 57\% after mounting the antenna. The computed reflectivities (see inset in Fig. \ref{fig:sim_farfield}.(a)) indicate the reflectivity decreases slightly with the antenna, so the outcoupled power should increase. This is true for the case when the antenna is perfectly aligned and attached to the laser, however after mounting, a gap of an estimated 25 \SI{}{\micro\metre} between the waveguide and antenna was observed. The measured antenna should therefore have a lower impact on the end facet reflectivity, as testified also by the lack of change in the laser threshold. Simulations predict that gaps in the order of a few tens of microns do not deteriorate the far-field. Results of a more detailed study can be found in the Supplementary Material.

\par
\begin{figure}
\includegraphics[width=0.45\textwidth]{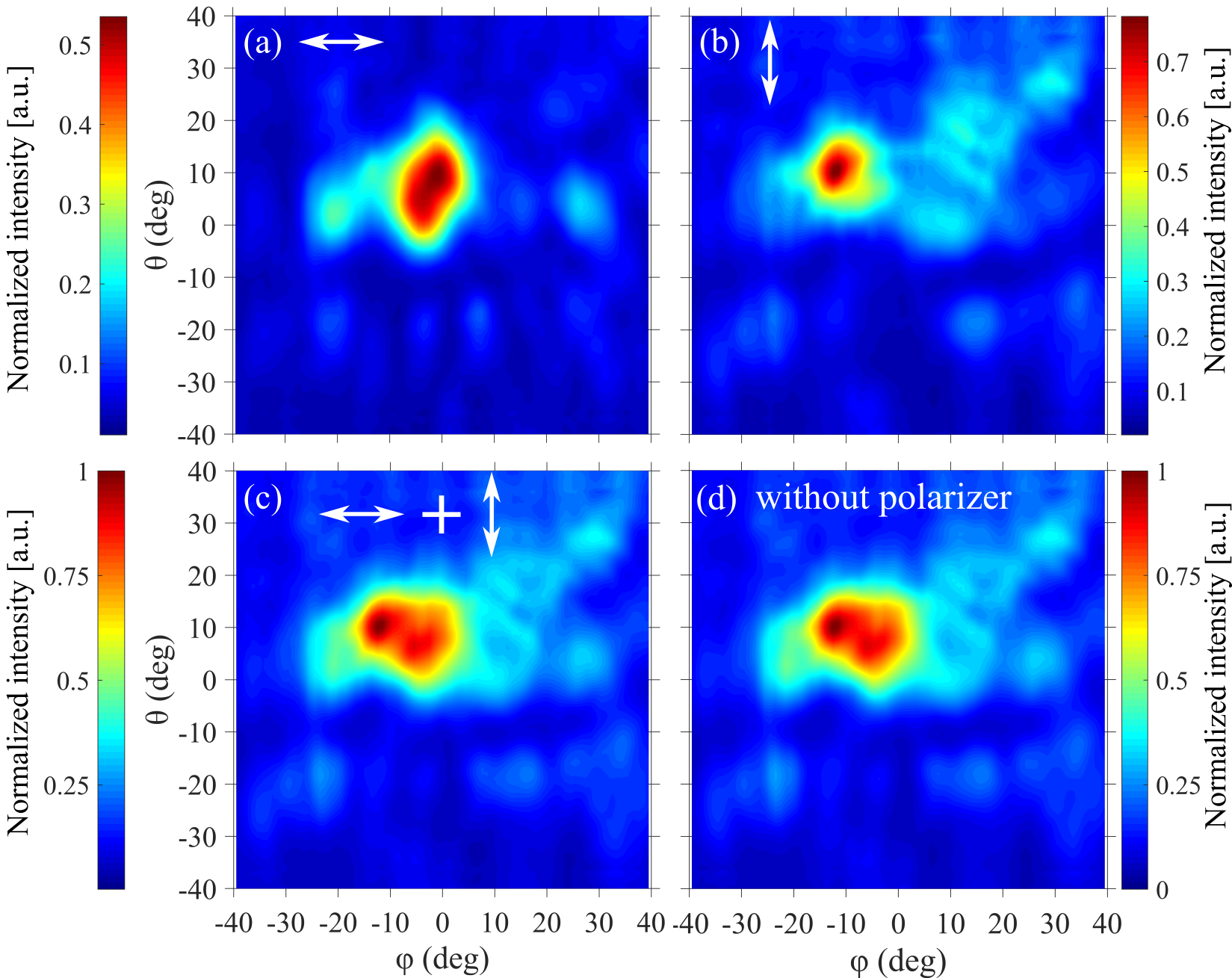}
\caption{\label{fig:polarization} Polarization-resolved far-field measurements using a wire-grid polarizer in front of the detector. (a) Measured horizontal component of the far-field with a FWHM of (15\degree$\times$21\degree). Intensity in this polarization component is detected only after mounting a Vivaldi antenna in front of the waveguide. (b) Measured vertical component of the far-field with a FWHM of (14\degree$\times$14\degree). (c) The sum of the measured horizontal and vertical far-field from (a) and (b) reproduces the initial far-field measurement without a polarizer, shown again in (d). The difference between these two plots does not exceed 4\% in any point. Intensities in (a)-(c) are normalized to the maximum value in (c), and (d) has the same range [0,1] for comparison.}
\end{figure}

To evaluate the polarization properties, we repeated the far-field measurements with a linear wire-grid polarizer placed in front of the detector. The polarizer was set for a maximum transmission of the vertical or horizontal component, respectively. We first did a reference measurement with a broadband double metal waveguide THz QCL, and the emission was, as expected,  purely vertically polarized. When measuring a device with the Vivaldi antenna, we observed a focused intensity in both the horizontal and the vertical component of the electric field, see Fig. \ref{fig:polarization}.(a, b). A sum of the two intensities in Fig. \ref{fig:polarization}.(c): $|E_{\mathrm{abs}}|^2 = |E_{\mathrm{hor}}|^2+|E_{\mathrm{ver}}|^2$ reconstructs the original far-field measurement, shown again in Fig. \ref{fig:polarization}.(d). The Vivaldi antenna acts as a wave retarder, rotating the polarization after the optical mode exits the waveguide end facet and follows the antenna's exponentially curved shape. This is predicted by simulations, which also give us an insight in the time-resolved propagation of light and its polarization. Moving through the antenna cross-section, there are regions where the electric field is elliptically or circularly polarized, as shown in the top right inset in Fig. \ref{fig:sim_farfield}.(a) (Multimedia view). In the far-field, the polarization is mainly linearly polarized, rotated by an angle from vertical towards horizontal polarization. This angle is frequency-dependent and is larger for lower frequency components.
\par
\begin{figure}
\includegraphics[width=0.45\textwidth]{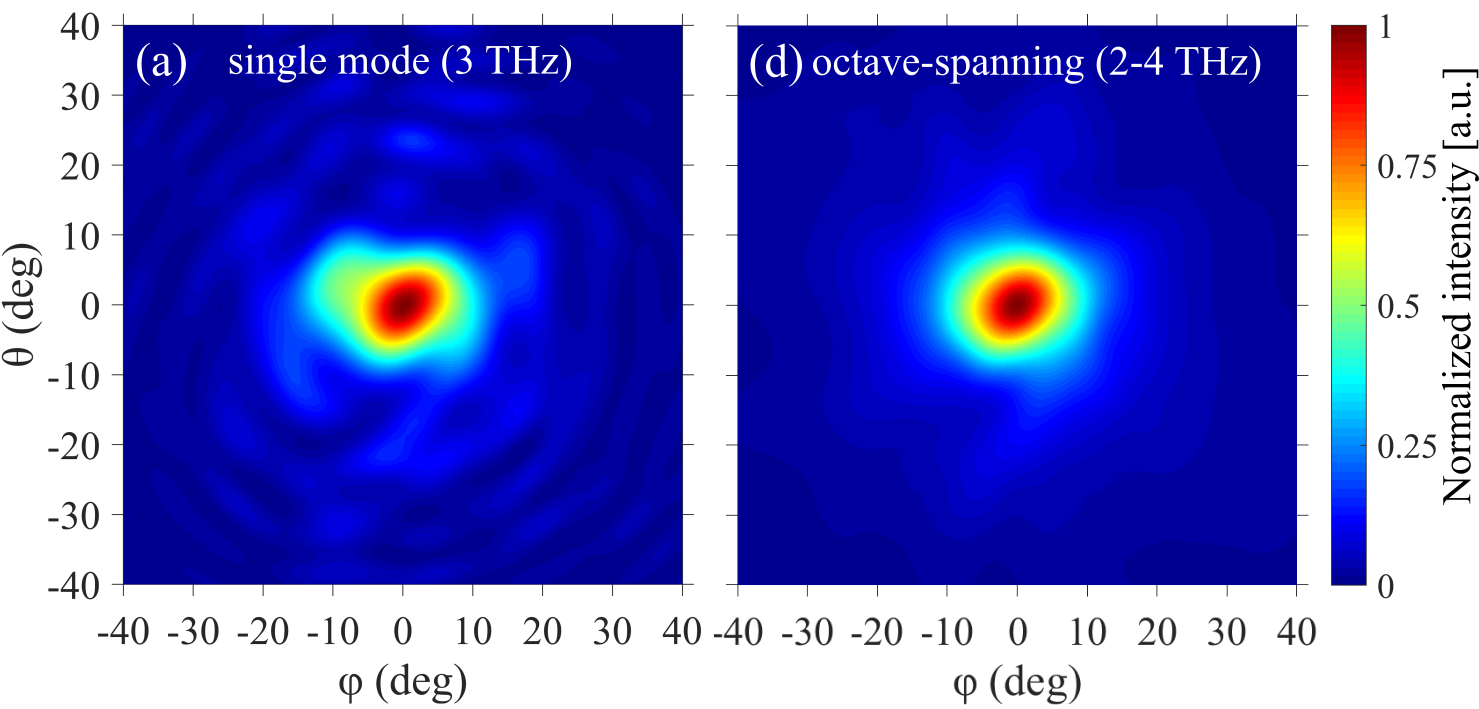}
\caption{\label{fig:farfield_spectrum} (a) Simulated far-field of a single-mode laser at 3 THz, FWHM (17\degree$\times$14\degree). (b) Simulated far-field of a broadband octave-spanning laser (2-4 THz), FWHM (16\degree$\times$14\degree). }
\end{figure}

 
We studied how the far-field changes depending on the spectrum. Simulations show that the Vivaldi antenna can be used for any shape of the emission spectrum within the designed spectral range between 1.5-4.5 THz, ranging from single mode to (multi)octave-spanning lasers. A comparison in Fig. \ref{fig:farfield_spectrum} shows that the FWHM beam widths of a single mode and an octave-spanning laser with the Vivaldi antenna differ by only around 1\degree. For broadband emission, the beam is more circular: at each single frequency, the beam is focused with a slightly elliptical shape, and with some intensity outside the main lobe. This ellipse is rotated by a different angle at each frequency, and the intensity outside the main lobe has a different shape. With broadband devices, the overlapping rotated elliptical beams produce a more circular pattern, while the intensities outside the main lobe get smeared out. 

\par
\begin{figure}
\includegraphics[width=0.4\textwidth]{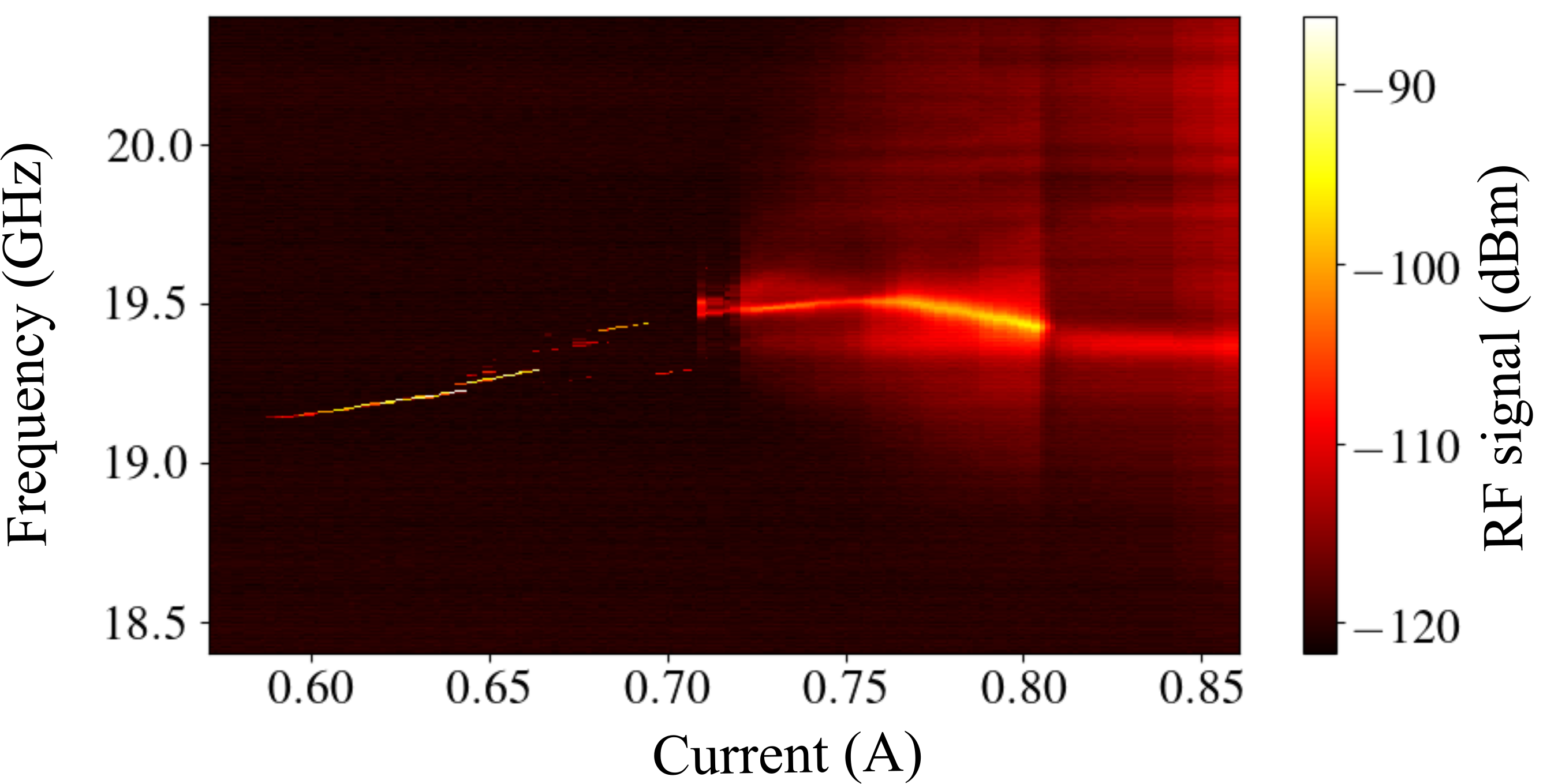}
\caption{\label{fig:beatnote} Beatnote map of a broadband THz QCL with a mounted Vivaldi antenna in pulsed mode (50 \SI{}{\micro s} pulses, 4 kHz at 15 K). A narrow beatnote until 700 mA is observed, indicating comb operation.}
\end{figure}

 
To verify that the Vivaldi antenna preserves the comb state, we measured the spectrum and beatnote at different applied biases in pulsed mode (50 \SI{}{\micro s} pulses, 4 kHz at 15 K). The resulting beatnote map is shown in Fig. \ref{fig:beatnote}. In a wide current  region until 700 mA a narrow beatnote is observed, which is a sign of comb operation  \cite{BurghoffNatPhot2014,Roesch2014}. The maximum spectral bandwidth of the laser in this range is 300 GHz. Similar results are observed in devices of the same epilayer, confirming the robustness of the comb state to the implementation of a Vivaldi antenna.

\par
We presented a broadband polarization-rotating antipodal Vivaldi antenna. When mounted in front of a double metal waveguide THz QCL, it produces a single-lobed far-field with a FWHM of (23\degree$\times$19\degree), a significant improvement compared to standard double metal waveguides. The excellent far-field properties are observed with broadband devices and agree well with numerical simulations, which also show that in the frequency range of interest (1.5-4.5 THz), the antenna works well with broadband as well as with single-mode lasers. The Vivaldi antenna does not have an effect on the laser's emission spectrum or current-voltage characteristics. The total collected power decreases by $\approx$20\%, but the power areal density increases by $\approx$57\% due to the more focused beam. Owing to its exponentially curved shape, the antenna acts as a wave retarder and rotates the polarization from vertical towards horizontal polarization by a frequency-dependent angle, advantageous for optical feedback reduction. There are also regions with elliptical or circular polarization. With a modified design of the antenna geometry (length and opening size), the emitted polarization at a specific frequency could be manipulated to produce a linearly polarized wave at a certain angle, or an elliptically/circularly polarized wave. The antenna could also be used as an input coupler which would transform the incident wave polarization to vertical polarization, perpendicular to the stack of quantum wells, increasing useful intensity of in-coupled radiation. When detecting the output radiation of a Vivaldi antenna with a polarization-sensitive detector, a complementary Vivaldi antenna could be placed in front of the detector to transform the polarization back to linear polarization, as initially emitted by the QCL.
Finally, the Vivaldi antenna is a planar device, and with suitable modifications and improvements in the fabrication, in principle it could be integrated with the THz QCL in the same  process flow. 
\section*{Supplementary material}
See supplementary material for a study of the influence of the air gap on the device performance in terms of the far-field, outcoupling efficiency and group delay dispersion.

\section*{Acknowledgments}
The authors acknowledge funding from the ERC grant  CHIC (724344). We thank the company TOPAS Advanced Polymers GmbH for free samples of TOPAS\textsuperscript{®} COC, and J. Andberger and F. Appugliese for technical help.

\bibliographystyle{apsrev4-1}

\bibliography{APL_Vivaldi_main}

\end{document}